\documentclass[aps,%
12pt,%
notitlepage,%
oneside,%
onecolumn,%
nobibnotes,%
nofootinbib,%
superscriptaddress,%
noshowpacs,%
centertags
]{revtex4} 

\usepackage[T2A]{fontenc}
\usepackage[utf8x]{inputenc}
\usepackage[russian,english]{babel}
\usepackage{graphicx}% Include figure files
\usepackage{dcolumn}% Align table columns on decimal point
\usepackage{tabularx}
\usepackage{bm}% bold math
\usepackage{amsmath}
\usepackage{booktabs}
\usepackage[flushleft]{threeparttable}
\usepackage[dvipsnames]{xcolor}
\usepackage{subcaption}
\usepackage{ragged2e}
\usepackage{hyperref}
\usepackage{cleveref}
\graphicspath{{pictures/}}

\begin{document}

\title{Infall onto the Protoplanetary Disk during the Gravitational Collapse of a Molecular Cloud}
\author{E.R. Redkin\(^{a,b,*}\) and E.I. Vorobyov\(^{a,c}\)\\
\textit{\(^{a}\) Research Institute of Physics, Southern Federal University, Rostov-on-Don, Russia}\\
\textit{\(^{b}\) Faculty of Physics, Southern Federal University, Rostov-on-Don, Russia}\\
\textit{\(^{c}\) Department of Astrophysics, University of Vienna, Vienna, Austria}\\
\textit{\(^{*}\)e-mail: {\rm redkin@sfedu.ru}}
}
\date{Received July 17, 2025; revised September 11, 2025; accepted October 13, 2025}

\begin{abstract}

\textbf{Abstract:}

The development of models for matter infall from a collapsing molecular cloud is an essential part of numerical studies on the formation and evolution of protoplanetary disks. In this article, the widely used Nakamoto and Nakagawa model is analyzed and modifications are proposed to complement the initial model. These improvements include calculation for the outer boundary of a molecular cloud and refinement of the initial density distribution within. Also, due to the finite size of a cloud, the approach for computing the rate of mass infall onto the protoplanetary disk during collapse is modified. The proposed enhancements are aimed at eliminating the key limitations of the initial model, but do not affect its advantages, such as simplicity in numerical implementation. Using the modified model, we investigate the duration of the embedded phase of the evolution of young stellar objects and compare the modeling results with observational data. The results indicate a shorter duration of the embedded phase of the evolution of young stellar objects, especially in star-forming regions with a high amplitude of the initial density perturbation of prestellar condensation.


\end{abstract}

\maketitle

\section*{1. INTRODUCTION}

According to modern concepts, stars form as a result of the gravitational collapse of cold (about 10-20 K) and slowly rotating molecular clouds. Not all of the matter from the collapsing molecular cloud falls directly onto the protostar: a significant portion of it, as a result of conservation of angular momentum, forms a rotating circumstellar disk \citep{Larson2003}. A further increase in the mass of the protostar occurs as a result of accretion of matter from the circumstellar disk. It is in such disks that the formation of planets subsequently occurs.

A key step in numerical modeling of the evolution of protoplanetary disks, as well as the formation of protostars and protoplanets, is development of models that describe the infall of matter from a collapsing cloud onto the disk (so-called infall models). There is a large class of one-dimensional models based on the theory of viscous evolution of accretion axisymmetric disks \citep{Weizsacker1943, LyndenBell&Pringle1974, Pringle1981}, which do not self-consistently calculate the transition from a collapsing molecular cloud to a ``protostar and protoplanetary disk'' system, but use simpler approaches to describe the rate of matter infall onto the protoplanetary disk.

One of the most frequently used approaches, related to calculating the trajectories of matter infall, was first described in \citep{Ulrich1976, Cassen&Moosman1981}. \citet{Cassen&Moosman1981} consider the collapse of an isothermal, spherically symmetric cloud rotating rigidly and assumes that matter falls onto an (infinitely) thin Keplerian disk along parabolic trajectories. It is noted that the falling gas at the point of contact with the disk has a lower azimuthal velocity than the azimuthal velocity of the gas in the Keplerian disk at a given distance from the star.

\citet{Nakamoto&Nakagawa1994} followed the approach of \citet{Cassen&Moosman1981} when modeling gravitational collapse, but they did not present a direct calculation of the trajectories of the matter infall. This is probably why these approaches are sometimes considered separately \citep{Lee2021}. However, the expression used in \citep{Nakamoto&Nakagawa1994} for the rate of matter infall per unit surface area of the protoplanetary disk $\dot{\Sigma}_{\rm infall}$ has the form
\begin{equation}
\dot{\Sigma}_{\rm infall} (R,t) = \frac{\dot{M}_{\rm c}}{4\pi R R_{\rm cf}} \left(1 - \frac{R}{R_{\rm cf}} \right)^{-\frac{1}{2}},
\label{SigmaInf_NN94}
\end{equation}
where $R$ is the radial distance (in the midplane of the disk); $R_{\rm cf}$ is the centrifugal radius, i.e., the distance in the disk at which the acceleration of the centrifugal force of matter infall will be at a given time $t$ equal to the acceleration of the force of gravitational attraction of the star; $\dot{M}_{\rm c}$ is the rate at which mass falls from the cloud onto the protostar and disk system. Equation (\ref{SigmaInf_NN94}) corresponds to what was obtained in \citep{Cassen&Moosman1981}. Therefore, matter infall in model \citep{Nakamoto&Nakagawa1994}, as in original study \citep{Cassen&Moosman1981}, occurs at a sub-Keplerian rate.

Traditionally, the viscous evolution of a protoplanetary disk is described by a one-dimensional equation from \citep{Pringle1981}, which, taking into account the additional source $\dot{\Sigma}_{\rm infall}$ due to the matter infall, takes the form
\begin{equation}
\frac{\partial \Sigma}{\partial t} = \frac{3}{R}\frac{\partial}{\partial R}\left(\sqrt{R} \frac{\partial}{\partial R} \left(\nu \Sigma \sqrt{R} \right)\right)  + \dot{\Sigma}_{\rm infall},
\label{Pringle_eq}
\end{equation}
where $\Sigma$ is the surface density of gas in the disk and $\nu$ is the kinematic viscosity. However, if the angular velocity of matter infall differs from Keplerian, then due to the conservation of angular momentum, the evolution equation in the form (\ref{Pringle_eq}) becomes inapplicable (see Appendix 1). The infall of matter at sub-Keplerian rates will lead to deceleration of matter in the disk and promote its movement in the radial direction toward the central star. As a result, the change in density in the disk at a given radial distance $R$ from the star is determined not only by viscous transfer and an increase due to the infall of matter, but also by a decrease in the angular momentum of the disk due to the infall of matter from the envelope.

In order to be able to use Eq. (\ref{Pringle_eq}) in its unaltered form, in \citep{Hueso&Guillot2005} a different expression for $\dot{\Sigma}_{\rm infall}$ was obtained from that used in \citep{Cassen&Moosman1981, Nakamoto&Nakagawa1994}, which implies equality of the angular momentum of the matter falling from the envelope and the matter in the disk at the point of their contact. However, as the authors of \citep{Visser2010} note, this approach leads to a discontinuity in the trajectories of the infalling matter, since this would mean that after falling, the matter instantly moves to an orbit of a smaller radius.

In fact, it is not at all necessary to abandon the approach of \citep{Nakamoto&Nakagawa1994}. Using the equation of continuity of mass and conservation of angular momentum, it can be shown that it is sufficient to add an additional term to Eq. (\ref{Pringle_eq}) (see Appendix 1), which takes into account the difference in the magnitude of the angular momentum of the matter infall and the angular momentum of the gas in the disk. The equation of viscous evolution of the disk then takes the form
\begin{equation}
\frac{\partial \Sigma}{\partial t} = \frac{3}{R}\frac{\partial}{\partial R}\left(\sqrt{R} \frac{\partial}{\partial R} \left(\nu \Sigma \sqrt{R} \right)\right)  + \dot{\Sigma}_{\rm infall} + \dot{\Sigma}_{\rm infall} \left(2 - 3 \sqrt{x} + \frac{x}{1 + \sqrt{x}} \right),
\label{Pringle_eq_modified}
\end{equation}
where $x = R/R_{\rm cf}$ is the ratio of the radial distance $R$ to the centrifugal radius $R_{\rm cf}$. For example, the authors of \citep{Liu2016} use the approach of \citep{Nakamoto&Nakagawa1994} and add the corresponding term to the equation of the evolution of the surface density, writing it in form (\ref{Pringle_eq_modified}).

There are also other models for matter infall from a collapsing cloud onto a protoplanetary disk. Thus, in \citep{Visser2009}, the boundary between the cloud and the disk is taken into account, the ratio of the scale height to the radial distance $H/R$ increases with the growth of $R$ (flared disk). The matter falls at the point where the trajectory of the falling matter first crosses this boundary. \citet{Schib2021}, in contrast, note that they do not calculate trajectories in their model, and instead use a normal (Gaussian) distribution as a source in the mass continuity equation, distributing the infalling material near a radial distance characterized by a specific angular momentum equal to the average specific angular momentum of the entire disk. The model described in works \citep{Takahashi2013, Kimura2016} also differs from those listed previously. As a density distribution in a molecular cloud, the authors use the Bonnor-Ebert sphere, and not the singular isothermal sphere, as in the model \citep{Nakamoto&Nakagawa1994}. Also in these works, the rate of matter falling from the shell onto the disk depends on the radial distance from the center of the cloud and is not constant over time.

It is worth noting that the original model \citep{Nakamoto&Nakagawa1994} has some shortcomings, which include the use of an idealized initial density distribution in the molecular cloud, the lack of calculation of the outer boundary of the cloud, and the constancy of the rate of mass loss from the cloud onto the central star and protoplanetary disk, which can have a significant impact on the evolution of the system \citep{Vorobyov&Basu2005}. However, the model \citep{Nakamoto&Nakagawa1994} is quite widely used \citep{Jin&Sui2010, Jin&Li2014J, Zhang&Jin2015, Xiao2016, Tang2020, Li2023, Yang2024}, and in this paper we will focus on proposing modifications to improve this approach.

\section*{2. ANALYSIS OF THE NAKAMOTO AND NAKAGAWA MODELS}

Let us analyze the approach to modeling the collapse of a molecular cloud and the formation of a protoplanetary disk described in \citep{Nakamoto&Nakagawa1994} in order to highlight the aspects that need to be modified. This approach considers a molecular cloud rotating rigidly with a constant angular velocity $\omega_{\rm c}$. Magnetic field effects are not taken into account. The temperature $T_{\rm c}$ of the molecular cloud is the same everywhere, does not change over time, and determines the isothermal sound speed $a$:
\begin{equation}
a = \sqrt{\frac{\mathcal{R} T_{\rm c}}{\mu}},
\label{a_isothermal}
\end{equation}
where $\mathcal{R}$ is the universal gas constant and $\mu$ is the average molar mass. All calculations in this article use the value $\mu = 2.33 \mathrm{~g} / \mathrm{mol}$.

It is assumed that the gas density distribution in the cloud is spherically symmetrical and at a distance of $r$ from the center of the cloud is determined by \citep{Shu1977}:
\begin{equation}
\rho(r) = \frac{A a^2}{4\pi Gr^2},
\label{rho_c_Shu77}
\end{equation}
where $G$ is the gravitational constant, $A$ is the density perturbation amplitude ($A \ge 2$), which is a positive uniform increase in density throughout the entire volume of the cloud. At $A = 2$, the gas density distribution is a marginally stable configuration, and at $A > 2$, the system leaves the state of equilibrium and begins to collapse, since at any radial distance, the local balance of gravity and pressure gradient is broken. A cloud with density distribution (\ref{rho_c_Shu77}) is called a singular isothermal sphere (SIS), since at $r \rightarrow 0$, $\rho \rightarrow \infty$.

The collapse occurs from the inside out \citep{Shu1977}, i.e., it starts at the center of the molecular cloud, and its front spreads spherically into the outer region at rate $a$. The rate of mass infall from the cloud onto the protostar and disk system is determined by the expression
\begin{equation}
\dot{M}_{\rm c} = m_0 \frac{a^3}{G},
\label{Mdotcore_Shu77}
\end{equation}
where the factor $m_0$ is taken as 0.975 if the amplitude $A$ in distribution (\ref{rho_c_Shu77}) is 2 \citep{Shu1977}. Since the cloud temperature is constant, the infall rate also does not change over time. The region in the disk within which the infall occurs is limited by the centrifugal radius
\begin{equation}
R_{\rm cf} = \frac{(l^2 \omega_{\rm c})^2}{GM(l)},
\label{R_cf}
\end{equation}
where $M(l)$ is the mass of the cloud enclosed within the radius $l$. This distance at time $t$ is determined by the expression
\begin{equation}
l = \frac{at}{2},
\label{R_t}
\end{equation}
where time $t$ is measured from the onset of gravitational collapse. For a singular isothermal sphere, this time also corresponds to the beginning of formation of the central condensation (protostar). Then the mass $M(l)$ can be calculated:
\begin{equation}
M(l) = \int\limits_0^l \rho(r) 4\pi r^2 dr = \int\limits_0^l \frac{A a^2}{4\pi Gr^2} 4\pi r^2 dr = \frac{Aa^2}{G} l.
\label{M_r_int_1}
\end{equation}
Taking into account equality (\ref{R_t}), by substituting Eq. (\ref{M_r_int_1}) into expression (\ref{R_cf}), we obtain an equation for the centrifugal radius:
\begin{equation}
R_{\rm cf}(t) = \frac{a\omega_{\rm c}^2t^3}{8A}.
\label{R_cf_NN94}
\end{equation}
The authors of \citep{Nakamoto&Nakagawa1994} emphasize that $R_{\rm cf} \propto t^3$, therefore, a significant increase in the centrifugal radius occurs at late times.

The effect of gravitational collapse of the protostellar cloud on the evolution of the protoplanetary disk in \citep{Nakamoto&Nakagawa1994} is taken into account by adding to the right-hand side of Eq. (\ref{Pringle_eq}) a second term characterizing the rate of matter infall per unit surface area of the protoplanetary disk $\dot{\Sigma}_{\rm infall}$, which is calculated according to Eq. (\ref{SigmaInf_NN94}). If it is assumed that matter infalls within the entire region $R<R_{\rm cf}$, then over the time interval $\Delta t$, the following equality is satisfied:
\begin{equation}
\int\limits_0^{R_{\rm cf}} \dot{\Sigma}_{\rm infall} (R,t) \Delta t \cdot 2\pi R dR = \dot{M}_{\rm c} \Delta t.
\label{CheckNorm_NN94}
\end{equation}
This equality means that the total mass of the matter infall during time $\Delta t$ is equal to the total mass distributed over the disk over the same time. Substituting Eq. (\ref{SigmaInf_NN94}) into relation (\ref{CheckNorm_NN94}) and integrating, it becomes clear that in \citep{Nakamoto&Nakagawa1994}, matter infall within the entire region $R<R_{\rm cf}$ is indeed assumed. Since the protoplanetary disk has an inner boundary $R_{\rm in}$ determined by the dynamic interaction of the disk with the magnetosphere of the protostar \citep{Gehrig2022}, infall occurs not only on the disk, but also on the protostar. Moreover, all the matter distributed in accordance with Eq. (\ref{SigmaInf_NN94}) in the range $0<R<R_{\rm in}$ is considered to have fallen onto the protostar. In this study, the value $R_{\rm in} = 0.05$ au was used in all models (median value according to \citet{Gehrig2021}).

The infall model based on the original approach of the authors of \citep{Nakamoto&Nakagawa1994} is called NN-0 below. The first row of Table~\ref{tab: table_models} shows the values of the cloud parameters $M_{\rm c}$, $\omega_{\rm c}$ и $T_{\rm c}$ used in the modeling, as well as the results of calculating the main physical quantities.

\begin{table}[!ht]
\begin{threeparttable}
    \centering
    \caption{Parameters of initial model (NN-0) and its modification (NN-0a)}
    \label{tab: table_models}
    \setlength{\tabcolsep}{6pt}
    \begin{tabularx}{\textwidth}{cccccccccccc}
    \toprule\noalign{\smallskip}
        \textbf{Model} & \textbf{$A$} & \textbf{$T_{\rm c}$} & \textbf{$\omega_{\rm c}$} & \textbf{$M_{\rm c}$} & \textbf{$r_{\rm out}$} & \textbf{$R_{\rm cf}(t_{\rm infall})$} & \textbf{$\dot{M}_{\rm c} (t=0)$} & \textbf{$M_{\rm infall}^{\rm star}$} & \textbf{$M_{\rm infall}^{\rm disk}$} & \textbf{$t_{\rm infall}$} \\ & & [$\rm K$] & [$\rm s^{-1}$] & [$\rm M_{\odot}$] & [$\rm au$] & [$\rm au$] & [$\rm M_{\odot}$ $\rm yr^{-1}$] & [$\rm M_{\odot}$] & [$\rm M_{\odot}$] & [$\rm kyr$] \\ \midrule\hline\noalign{\smallskip}
        \textbf{NN-0} & 2 & 15 & 2.8$\times$10$^{-14}$ & 1 & -- & 103 & 2.85$\times$10$^{-6}$ & 0.20 & 0.80 & 350.8 \\ \midrule
        \textbf{NN-0a} & 2 & 15 & 2.8$\times$10$^{-14}$ & 1 & 8337 & 103 & 2.85$\times$10$^{-6}$ & 0.20 & 0.80 & 350.8 \\  \bottomrule
    \end{tabularx}%
    \begin{tablenotes}
      \small
      \item 
        $A$ density perturbation amplitude \citep{Shu1977}; $T_{\rm c}$ molecular cloud temperature \citep{Jijina1999}; $\omega_{\rm c}$ angular velocity of molecular cloud \citep{Goodman1993, Caselli2002}; $M_{\rm c}$ molecular cloud mass; $r_{\rm out}$ molecular cloud radius; $R_{\rm cf}(t_{\rm infall})$ centrifugal radius at end of collapse; $\dot{M}_{\rm c} (t=0)$ initial rate of mass infall from collapsing cloud; $M_{\rm infall}^{\rm star}$ mass accreted onto star during collapse; $M_{\rm infall}^{\rm disk}$ mass accreted onto disk during collapse; $t_{\rm infall}$ time for entire cloud mass $M_{\rm c}$ to accrete onto star and disk.
    \end{tablenotes}
\end{threeparttable}
\end{table}

Taking into account that the rate of matter infall from the cloud $\dot{M}_{\rm c}$ does not change over time, it is possible to find the time of matter infall:
\begin{equation}
t_{\rm infall} = \frac{M_{\rm c}}{\dot{M}_{\rm c}}.
\label{t_inf}
\end{equation}
It is noteworthy that the size of the cloud in the initial model \citep{Nakamoto&Nakagawa1994} is not limited in any way, and the collapse continues until a mass equal to a given value $M_{\rm c}$ falls out.

\begin{figure}[!ht]
    \centering
    \includegraphics[scale=1.2]{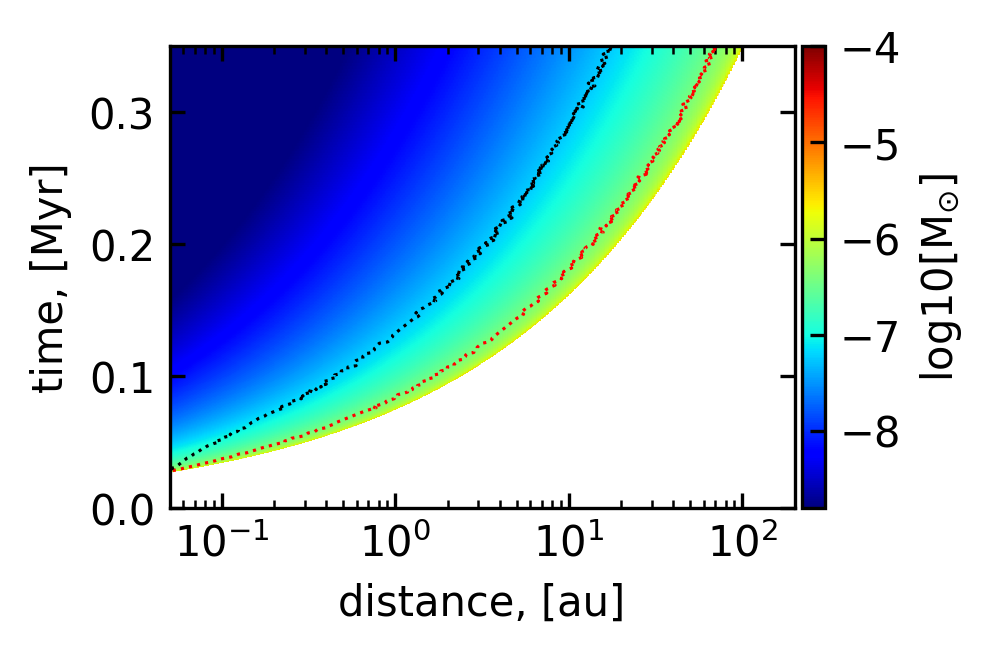}
    \caption{Distribution of mass (shown in color) of matter infall across disk in radial direction (along horizontal axis) and its change over time (along vertical axis). Boundary of colored region shows change in centrifugal radius of disk over time. Black and red dashed lines delineate boundaries of regions outside of which \(90\%\) and \(50\%\) of mass that fell onto disk is contained, respectively. Calculations were performed using NN-0 model.}
    \label{fig:Minf_NN94}
\end{figure}

The distribution of the fallen mass over the disk in the NN-0 model is shown in Fig.~\ref{fig:Minf_NN94}. The mass of the matter that fell during the time $\Delta t$ into a cell with area $S_{\rm cell}$ with the center located at distance of $R$ from the star is calculated by the formula
\begin{equation}
M_{\rm infall} (R,t) = \dot{\Sigma}_{\rm infall} (R,t) \Delta t \cdot S_{\rm cell}.
\label{Mass_infall}
\end{equation}
The nonuniformity of the distribution of the fallen mass over the disk is related, first, to the nonlinear increase in the area of the cells approaching the outer edge of the disk, and, second, with the nature of the dependence of $\dot{\Sigma}_{\rm infall}$ on distance $R$. From formula (\ref{SigmaInf_NN94}), it is evident that at $R \ll R_{\rm cf}$, the rate of matter infall per unit surface area of the disk decreases in inverse proportion to $R$, but as $R$ approaches $R_{\rm cf}$, the difference in brackets tends to zero and $\dot{\Sigma}_{\rm infall}$ increases significantly. Since the matter falls only within the centrifugal radius, the boundary of the colored region shows a change $R_{\rm cf}$ over time. The matter infall occurs only onto the star and the colored region in the figure does not appear until the centrifugal radius reaches the inner boundary of the disk ($R_{\rm cf} < R_{\rm in}$), which occurs at time  $t \approx 0.028$ Myr. At the same time, by the end of the collapse, as can be seen from Table~\ref{tab: table_models}, about \(20\%\) of the mass of the initial cloud passes directly to the star, while the remaining mass falls onto the disk. Thus, model \citep{Nakamoto&Nakagawa1994} represents a case of so-called disk-mediated protostellar accretion, in which most of the final mass of the star is accreted from the disk rather than directly from the collapsing cloud.

The considered NN-0 model is simple to use and implement numerically and therefore finds wide application in one-dimensional models of the viscous evolution of protoplanetary disks. However, one cannot help but note the weaknesses of the NN-0 model.

\begin{enumerate}
    \item The density distribution (\ref{rho_c_Shu77}) used in the model corresponds to SIS and is a mathematical idealization. However, when considering the central regions, this approximation cannot describe the real physical conditions in the cloud.
    \item The cloud size in this model is not limited in any way, which makes it impossible to estimate some key parameters of protostellar clouds, such as $\beta$ the ratio of the rotational energy to gravitational energy.
    \item The rate of matter falling out of the cloud does not change over time and simply ``turns off'' when a given amount of mass has fallen out of the cloud. This drawback becomes particularly relevant if a cloud size limit is introduced, which will be discussed in more detail in Subsection 3.3.
\end{enumerate}
This study is aimed at eliminating the above-mentioned shortcomings of the model \citep{Nakamoto&Nakagawa1994} and at comparing the modified model with observations in the context of the duration of the embedded stage of the evolution of young stellar objects.

\section*{3. RESULTS}

In Subsections 3.1 -- 3.3, modifications that could contribute to improving the model \citep{Nakamoto&Nakagawa1994} are presented and analyzed. Subsection 3.4 also presents the results of calculations using models that include the described modifications. In Subsection 3.5, the special case of disk-mediated stellar accretion is considered separately.

\subsection*{3.1. Cloud Size Limitation}

First, we supplement the model by introducing the calculation of the cloud boundary. For the distribution corresponding to SIS, the presence of a boundary is not assumed, but we assume that it is located at such a distance from the center of the system $r_{\rm out}$ that the mass inside it is $M(r_{\rm out}) = M_{\rm c}$. Using Eq. (\ref{M_r_int_1}), we obtain
\begin{equation}
r_{\rm out} = \frac{GM_{\rm c}}{Aa^2}.
\label{R_cloud_NN94}
\end{equation}
The introduction of this modification alone, of course, does not have any effect on the simulation results, but it allows us to calculate the ratio of rotational $E_{\rm rot}$ to gravitational $E_{\rm grav}$ energy for such a configuration. It should be noted that it is this ratio, and not the value $\omega_{\rm c}$, that largely determines the properties of protoplanetary disks formed as a result of the collapse of molecular clouds \citep{Vorobyov2012,Meyer2021}.

The rotational energy, taking into account distribution (\ref{rho_c_Shu77}), is determined as follows:
\begin{equation}
E_{\rm rot} = \frac{1}{2} \int\limits_0^{r_{\rm out}}{\rho(r) \omega_{\rm c}^2 r^2 \cdot 4 \pi r^2 dr} = \frac{A a^2 \omega_{\rm c}^2 r_{\rm out}^3}{6 G}.
\label{Erot}
\end{equation}
The expression for gravitational energy can also be obtained analytically taking into account distribution (\ref{rho_c_Shu77}) and Eq. (\ref{M_r_int_1}):
\begin{equation}
E_{\rm grav} = - \int\limits_0^{r_{\rm out}}{\rho(r) \cdot\frac{ G M(r)}{r^2} \cdot r \cdot 4 \pi r^2 dr} = -\frac{A^2 a^4 r_{\rm out}}{G}.
\label{Egrav}
\end{equation}
Let us find $\beta$ and, using relations (\ref{a_isothermal}) and (\ref{R_cloud_NN94}), show that it depends only on the cloud parameters $A$, $M_{\rm c}$, $\omega_{\rm c}$, and $T_{\rm c}$:
\begin{equation}
\beta = \frac{E_{\rm rot}}{|E_{\rm grav}|}=\frac{1}{6A} \left(\frac{\omega_{\rm c} r_{\rm out}}{a} \right)^2 = \frac{1}{6} \frac{G^2 \mu^3}{\mathcal{R}^3} \frac{(M_{\rm c} \omega_{\rm c})^2}{(AT_{\rm c})^3}.
\label{beta}
\end{equation}
The NN-0 model with calculation of the outer boundary of the disk is further called NN-0a. Using the cloud parameters from Table~\ref{tab: table_models}, we obtain for this model $r_{\rm out}$ = 8337 au and $\beta$ = 0.19$\%$. Note that this value $\beta$ falls within the range of values characteristic of prestellar molecular clouds \citep{Caselli2002}. Thus, this modification allows us to set more realistic initial conditions and directly correlate the properties of protoplanetary disks with key parameters of their parent clouds.

\subsection*{3.2. Modification of the Density Distribution $\rho(r)$}

The shortcomings of the NN-0 model noted at the end of Section 2 are primarily related to the fact that it is based on a mathematically convenient, but idealized, cloud density distribution (\ref{rho_c_Shu77}). We propose using an alternative density distribution characterized by a central plateau and described in \citep{Larson1969, Penston1969}:
\begin{equation}
\rho(r) \equiv \rho_{\rm LP}  = \frac{\rho_0}{1 + \frac{r^2}{r_0^2}},
\label{rho_c_LP69}
\end{equation}
where $\rho_0$ is the central density, which is related to the central concentration $n_0$ by the formula $\rho_0 = n_0 m_{\rm H} \mu$ where $m_{\rm H}$ is the mass of the hydrogen atom. The value of $n_0$ is taken as \(10^{10}~\mathrm{cm}^{-3}\), which corresponds to the concentration in the center of the cloud before the formation of the first hydrostatic core \citep{Vaytet2018}. The characteristic spatial scale of the central plateau $r_0$ is compared with the Jeans wavelength and is determined by the expression
\begin{equation}
r_0 = \frac{ka}{\sqrt{\pi G \rho_0}}.
\label{r_0_LP}
\end{equation}
The coefficient $k$ entering into equality (\ref{r_0_LP}) regulates the size of the central plateau and is calculated from the following considerations. As $\rho_0$ tends to infinity, the density distribution (\ref{rho_c_LP69}) (hereinafter $\rho_{\rm LP}$) becomes similar to the profile $\rho_{\rm SIS}$ represented by expression (\ref{rho_c_Shu77}). Let us show this explicitly using equality (\ref{r_0_LP}):
\begin{equation}
\rho_{\rm LP}^{\rm asm} = \lim_{\rho_0 \rightarrow \infty} \frac{\rho_0}{\rho_0 \left(\frac{1}{\rho_0} + \frac{\pi G r^2}{k^2 a^2} \right)} = \frac{k^2a^2}{\pi G r^2}.
\label{rhoLP_inf}
\end{equation}
Comparison of this expression with distribution (\ref{rho_c_Shu77}) allows us to conclude that when choosing the coefficient $k$ in accordance with the formula
\begin{equation}
k = \frac{\sqrt{A}}{2},
\label{factor_k}
\end{equation}
the asymptotic profile $\rho_{\rm LP}^{\rm asm}$ will coincide with the profile $\rho_{\rm SIS}$ for a given value of the density perturbation amplitude $A$.

In the limiting case of collapse of a molecular cloud with density distribution (\ref{rho_c_LP69}), the value of $A$ will be 8.8 \citep{Penston1969}, which means an increase in density by 4.4 times compared to the unperturbed configuration (at $A = 2$), but to bring the system out of equilibrium, an increase in density by \(10-30\%\) is sufficient \citep{Vorobyov2024}.

\begin{figure}[!ht]
    \centering
    \includegraphics[scale=0.6]{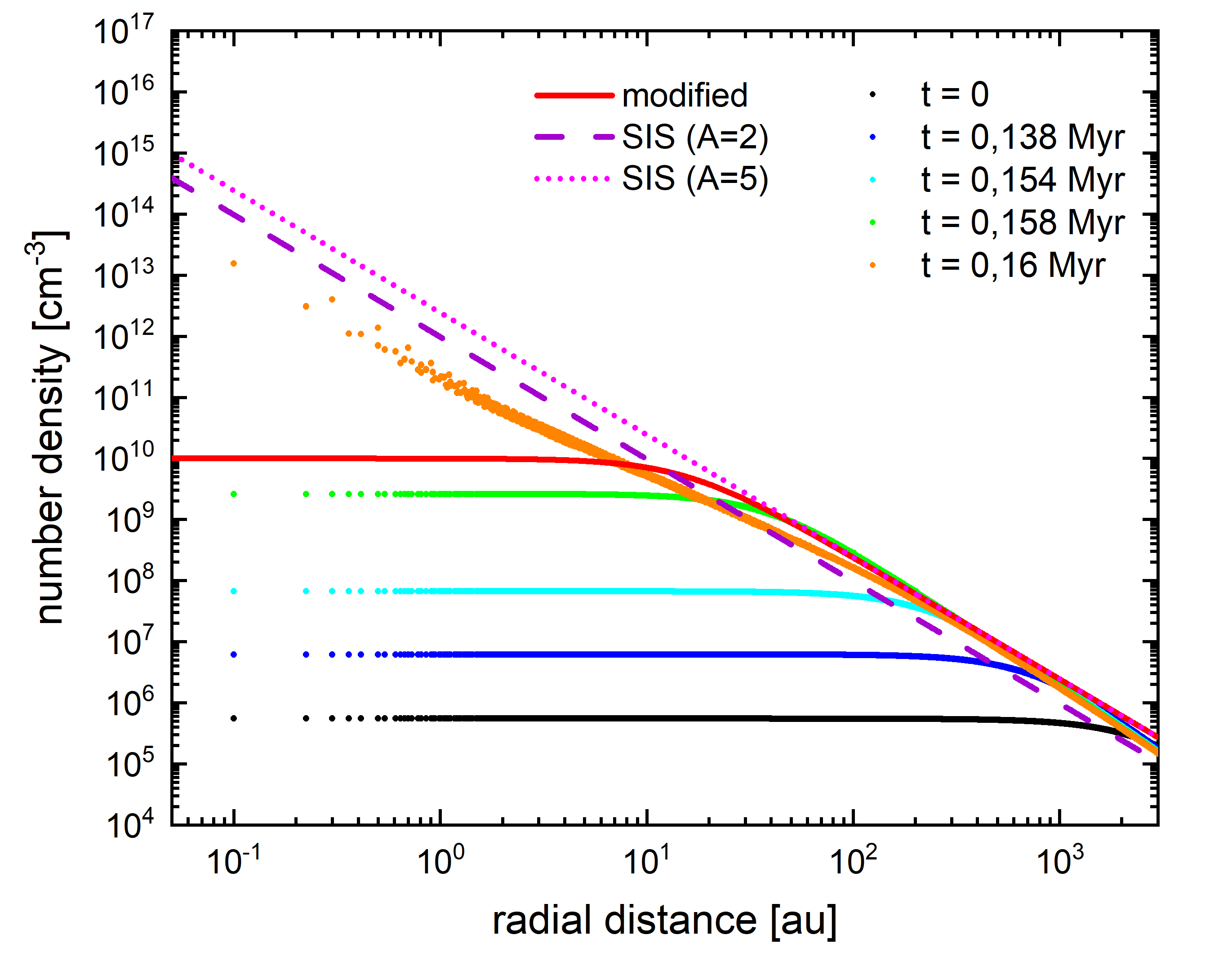}
    \caption{Gas concentration profiles in molecular cloud at $T_{\rm c}$ = 10~K. Dots indicate profiles constructed for different moments of time based on three-dimensional modeling data \citep{Vorobyov2024}. Constant concentration profile constructed using distribution (\ref{rho_c_LP69}) at $T_{\rm c}$ = 10~K and $A$ = 5 is shown by red solid line. Profiles corresponding to singular isothermal sphere with distribution (\ref{rho_c_Shu77}) for $A =$ 2 and 5 are plotted as dotted and dashed lines, respectively.}
    \label{fig:number_dens}
\end{figure}

Fig.~\ref{fig:number_dens} shows the gas concentration profiles constructed from the data of three-dimensional numerical hydrodynamic modeling of a collapsing cloud \citep{Vorobyov2024} and the concentration profile constructed using distribution (\ref{rho_c_LP69}) at $T_{\rm c}$ = 10~K и $A$ = 5. In both cases, the gas density before formation of the protostar ($t < 0.16$~Myr) shows a characteristic behavior in which the radial profile has no singularity and the density reaches a plateau, the size of which decreases as the collapse progresses. Note that the type of profile (\ref{rho_c_LP69}) corresponds to a Bonnor-Ebert sphere \citep{Bonnor1956, Ebert1955}, and at large distances from the center of the cloud, it coincides with distribution (\ref{rho_c_Shu77}) at $A$ = 5, which lies 2.5 times higher than the SIS at $A$ = 2 and almost completely coincides with the resulting numerical hydrodynamic modeling profile.

When the cloud density distribution described by Eq. (\ref{rho_c_Shu77}) changes to distribution (\ref{rho_c_LP69}), expression (\ref{R_cf_NN94}) for the centrifugal radius $R_{\rm cf}$ also changes. To derive the new expression $R_{\rm cf}$, we first obtain the relation for the mass of a cloud enclosed within a sphere of radius $l$:
\begin{equation}
M(l) = \int\limits_0^l \rho(r) 4\pi r^2 dr = 4 \pi \rho_0 \int\limits_0^R \frac{r^2}{1 + \frac{r^2}{r_0^2}} dr = 4 \pi \rho_0 r_0^3 \left( \frac{l}{r_0} - \arctan{\frac{l}{r_0}} \right).
\label{M_r_int_2}
\end{equation}
Taking into account equalities (\ref{R_t}) and (\ref{M_r_int_2}), the expression for the centrifugal radius (\ref{R_cf}) takes the form
\begin{equation}
R_{\rm cf}(t) = \frac{1}{64} \frac{\omega_{\rm c}^2 a^4 t^4}{\pi G \rho_0 r_0^3 \left( \frac{at}{2r_0} - \arctan{\frac{at}{2r_0}} \right)}.
\label{R_cf_modified}
\end{equation}
Taking into account expressions (\ref{r_0_LP}) and (\ref{factor_k}), in the limit $\rho_0 \rightarrow \infty$ this equation becomes Eq. (\ref{R_cf_NN94}), which characterizes the change in the centrifugal radius over time in the unmodified model.

Since the expression for the mass has changed, the new cloud radius $r_{\rm out}$ cannot be found using formula (\ref{R_cloud_NN94}). However, it can be found numerically (by any method of finding the roots of an equation) using
expression (\ref{M_r_int_2}), provided $M(r_{\rm out})=M_{\rm c}$ .

\subsection*{3.3. Expression for the Rate of Matter Infall $\dot{M}_{\rm c}$}

The modifications made to the model \citep{Nakamoto&Nakagawa1994} in Subsections 3.1 and 3.2 make it necessary to reconsider the approach to calculating the rate of mass infall from the collapsing cloud $\dot{M}_{\rm c}$.

First, the factor $m_0$ included in the equation for the infall rate is related to the amplitude of the density perturbation $A$ since the infall rate depends on how strongly the initial configuration deviates from the equilibrium position. A detailed analysis of the relationship between these parameters and their numerical values is given in \citep{Shu1977}. The table presented in this paper is limited to the value $A = 4$ however, based on the available values $m_0$ and $A$ the relationship between them can be approximated by the following polynomial:
\begin{equation}
m_0(A) = -4.7568 + 3.8059 \cdot A - 0.6292 \cdot A^2 + 0.0811 \cdot A^3.
\label{polynom_ShuTable}
\end{equation}
It is noteworthy that to calculate the coefficient $m_0$ we can also use the formula from \citep{Kratter2010}. However, as Appendix 2 shows, the values of $m_0$ for $2 < A \lesssim 5$ found in this way are somewhat overestimated compared to those presented in \citep{Shu1977}, which gives a higher rate of matter infall. Therefore, for relatively small $A$ it is preferable to use the values from initial study \citep{Shu1977} or polynomial (\ref{polynom_ShuTable}).

Second, Eq. (\ref{Mdotcore_Shu77}) for the rate of matter infall from the cloud ceases to be relevant, since it implies a constant $\dot{M}_{\rm c}$ which is only possible if the size of the cloud is not limited. Therefore, the initial model does not take into account the limited mass reservoir. The collapse ends when a given amount of matter falls out of the cloud, according to formula (\ref{t_inf}). If $M_{\rm c}$ is increased or decreased, $t_{\rm infall}$ will change proportionally, but the rate of matter infall will remain the same.

As discussed in \citep{Vorobyov&Basu2005}, the rate of mass infall $\dot{M}_{\rm c}$ from the cloud onto the disk ceases to be quasi-constant and begins to decay when the rarefaction wave coming from the outer boundary of the cloud collides with the rarefaction wave moving from the inside out. This happens at time
\begin{equation}
t_{\rm d} = \frac{r_{\rm out}}{2a}.
\label{t_damp}
\end{equation}
Then, the following modification for the infall rate can be proposed (hereinafter, modification M1):
\begin{equation}
    \dot{M}_{\rm M1} = \left\{
    \begin{alignedat}{2} 
        m_0\frac{a^3}{G}, & \quad t < t_{\rm d}, \\
        m_0\frac{a^3}{G} \cdot e^{\frac{t_{\rm d}-t}{t_{\rm d}}}, & \quad t \ge t_{\rm d}.
    \end{alignedat}
    \right.
\label{M_c_dot_M1}
\end{equation}

\begin{figure}[!ht]
    \centering
    \includegraphics[scale=0.5]{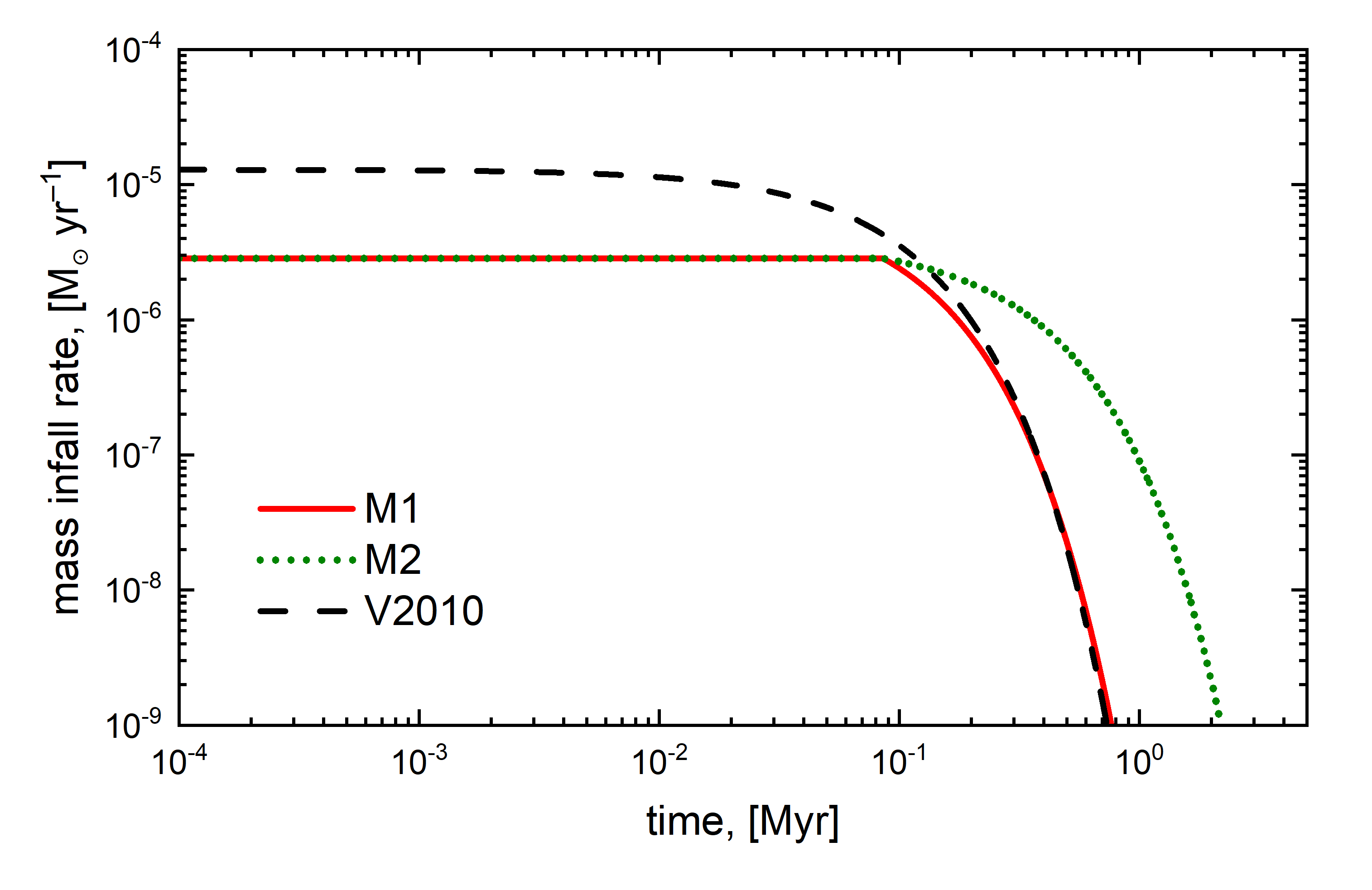}
    \caption{Changes in rate of matter infall over time using various modifications. Solid red line is M1 modification; dashed green line, M2 modification. Dashed black line shows infall rate as calculated by V2010. Same cloud parameters were used in all cases: $M_{\rm c}$ = 1 $M_{\rm \odot}$, $r_{\rm out}$ = 8337~au, $T_{\rm c}$ = 15~K и $\omega_{\rm c}$ = 2.8$\times$10$^{-14}$ s$^{-1}$. To enable comparison with V2010 calculations, density distribution (\ref{rho_c_Shu77}) characteristic of SIS for $A$ is used.}
    \label{fig:Mcoredot_M1_M2_V2010}
\end{figure}

For comparison, we take the expression for the rate of matter infall proposed in \citep{Vorobyov2010} based on numerical hydrodynamic modeling of the embedded stage of star formation (hereinafter V2010):
\begin{equation}
\dot{M}_{\rm V2010}=\frac{3M_{\rm c}}{t_{\rm ff}} \cdot e^{-\frac{3t}{t_{\rm ff}}}.
\label{M_c_dot_V2010}
\end{equation}
The time of free fall $t_{\rm ff}$ of matter at the cloud boundary, where the volume density $\rho(r_{\rm out})=\rho_{\rm out}$ enters into Eq. (\ref{M_c_dot_V2010}), is determined as follows:
\begin{equation}
t_{\rm ff} =\sqrt{\frac{3\pi}{32G\rho_{\rm out}}}.
\label{t_ff}
\end{equation}
Although mathematically the decay in Eq. (\ref{M_c_dot_V2010}) begins at time $t$ = 0 in reality it is initially so insignificant that it almost completely replicates the nature of the dependence in the modification M1 with an initial constant rate (plateau), as shown in Fig.~\ref{fig:Mcoredot_M1_M2_V2010}. However, this dependence is difficult to control, since the form of the expression included in it for $t_{\rm ff}$ depends on the chosen volume density distribution in the cloud.

At late times, as Fig.~\ref{fig:Mcoredot_M1_M2_V2010} shows, the character of decay of $\dot{M}_{\rm c}$ using modification M1 (Eq. (\ref{M_c_dot_M1})) corresponds to that in V2010. However, expression (\ref{M_c_dot_M1}) has a significant drawback: it does not take into account the initial mass of the cloud. The total mass that could fall onto the protostar and disk system over time can be found by integrating:
\begin{equation}
M_{\rm infall} = \int\limits_0^\infty{\dot{M}_{\rm c}(t)dt} = \int\limits_0^{t_{\rm d}}{m_0\frac{a^3}{G}dt} + \int\limits_{t_{\rm d}}^\infty{m_0\frac{a^3}{G}\cdot e^{\frac{t_{\rm d}-t}{t_{\rm d}}}dt} = m_0\frac{a^3}{G} t_{\rm d} + m_0\frac{a^3}{G} t_{\rm d},
\label{integral_M1}
\end{equation}
where the first integral shows the mass lost from the beginning of the collapse until the moment $t_{\rm d}$ when the decay begins, and the second one shows the mass lost after that. However, integration gives the same result for both the first and second terms, therefore, after the onset of decay, more mass cannot fall out than has already fallen out by the time $t_{\rm d}$.

Thus, Eq. (\ref{M_c_dot_M1}) has limitations. If the decay begins before the $0.5 M_{\rm c}$ infall, then some of the remaining mass will be ``locked'' in the cloud, since it will not have time to fall onto the disk and protostar before the rate $\dot{M}_{\rm c}$ decreases to very low values due to exponential decay. Of course, the expression for the infall rate must be such that for \textit{any} initial cloud parameters the following holds:
\begin{equation}
M_{\rm infall} = \int\limits_0^\infty{\dot{M}_{\rm c}(t)dt} = M_{\rm c}.
\label{integral_Minf}
\end{equation}
Taking these shortcomings into account, the following approach to calculating the rate of mass infall is proposed (modification M2):
\begin{equation}
    \dot{M}_{\rm M2} = \left\{
    \begin{alignedat}{2} 
        m_0\frac{a^3}{G}, & \quad t < t_{\rm d}, \\
        m_0\frac{a^3}{G} \cdot e^{\frac{t_{\rm d}-t}{t_{\rm fin}-t_{\rm d}}}, & \quad t \ge t_{\rm d},
    \end{alignedat}
    \right.
\label{M_c_dot_M2}
\end{equation}
where $t_{\rm fin}$ is the time it would take for all the matter to pass from the cloud into the protostar and disk system at a constant rate of infall, as it was at the initial time:
\begin{equation}
t_{\rm fin} =\frac{M_{\rm c}}{\dot{M}_{\rm c}(t=0)}.
\label{t_fin}
\end{equation}
Expression (\ref{M_c_dot_M2}) for any $M_{\rm c}$, $T_{\rm c}$ and $\omega_{\rm c}$ satisfies condition (\ref{integral_Minf}). It is this approach that is used below in the models NN-1 -- NN-7. The time of completion of matter infall $t_{\rm infall}$ can no longer be calculated using formula (\ref{t_inf}) and, since Eq. (\ref{M_c_dot_M2}) includes a decaying exponential, it should be determined based on the fraction of mass remaining in the cloud, which will be considered sufficiently small to indicate the end of collapse. In the models used in this study, the threshold value was chosen as $M_{\rm c}(t_{\rm infall}) = 10^{-3} M_{\rm c}(t=0)$ which corresponds to the loss of \(99.9\%\) of the initial cloud mass.

\subsection*{3.4. Duration of the Embedded Phase}

Let us analyze the results of calculations using models NN-1 -- NN-7 (see Table~\ref{tab: table_NNmodels}), including all previously proposed modifications: density distribution (\ref{rho_c_LP69}) and calculation of the outer boundary of the cloud and calculation of the rate of mass infall using Eq. (\ref{M_c_dot_M2}). The cloud parameters in all models are selected in such a way that the ratio of rotational energy to gravitational energy is $\beta \approx 0.2\%$ as in the NN-0a model.

\begin{table}[!ht]
\begin{threeparttable}
    \centering
    \caption{Parameters of models used and modeling results}
    \label{tab: table_NNmodels}
    \setlength{\tabcolsep}{4.75pt}
    \begin{tabularx}{\textwidth}{ccccccccccccc}
    \toprule\noalign{\smallskip}
        \textbf{Model} & \textbf{$A$} & \textbf{$T_{\rm c}$} & \textbf{$\omega_{\rm c}$} & \textbf{$M_{\rm c}$} & \textbf{$r_{\rm out}$} & \textbf{$R_{\rm cf}(t_{\rm infall})$} & \textbf{$\dot{M}_{\rm c} (t=0)$} & \textbf{$t_{\rm C0}$} & \textbf{$t_{\rm CI}$} & \textbf{$t_{\rm infall}$} & \textbf{$t_{\rm d}$} \\ & & [$\rm K$] & [$\rm s^{-1}$] & [$\rm M_{\odot}$] & [$\rm au$] & [$\rm au$] & [$\rm M_{\odot}$ $\rm yr^{-1}$] & [$\rm kyr$] & [$\rm kyr$] & [$\rm kyr$] & [$\rm kyr$] \\ \midrule\hline\noalign{\smallskip}
        \textbf{NN-1} & 3.0 & 10 & 2.8$\times$10$^{-14}$ & 1 & 8356 & 392 & 5.1$\times$10$^{-6}$ & 98.6 & 247.1 & 671 & 105  \\ \midrule
        \textbf{NN-2} & 4.0 & 10 & 4.4$\times$10$^{-14}$ & 1 & 6275 & 47 & 8.9$\times$10$^{-6}$ & 56.2 & 115.6 & 270 & 79  \\ \midrule
        \textbf{NN-3} & 5.0 & 10 & 6.0$\times$10$^{-14}$ & 1 & 5027 & 4.5 & 1.4$\times$10$^{-5}$ & 36.2 & 65.4 & 108 & 63  \\ \midrule
        \textbf{NN-4} & 3.0 & 15 & 5.1$\times$10$^{-14}$ & 1 & 5582 & 255 & 9.3$\times$10$^{-6}$ & 53.6 & 134.3 & 365 & 57  \\ \midrule
        \textbf{NN-5} & 4.0 & 15 & 7.8$\times$10$^{-14}$ & 1 & 4195 & 29 & 1.6$\times$10$^{-5}$ & 30.6 & 62.8 & 146 & 43  \\ \midrule
        \textbf{NN-6} & 3.0 & 7 & 1.7$\times$10$^{-14}$ & 1 & 11926 & 568 & 3.0$\times$10$^{-6}$ & 168.3 & 422.2 & 1147 & 179  \\ \midrule
        \textbf{NN-7} & 2.2 & 10 & 1.8$\times$10$^{-14}$ & 1 & 11385 & 6415 & 2.3$\times$10$^{-6}$ & 230.9 & 705.8 & 2064 & 143  \\ \bottomrule
    \end{tabularx}%
    \begin{tablenotes}
      \small
      \item 
      $A$ density perturbation amplitude; $T_{\rm c}$ is molecular cloud temperature; $\omega_{\rm c}$ angular velocity of molecular cloud; $M_{\rm c}$ molecular cloud mass; $r_{\rm out}$ molecular cloud radius; $R_{\rm cf}(t_{\rm infall})$ centrifugal radius at end of collapse; $\dot{M}_{\rm c} (t=0)$ initial the rate of mass infall from the collapsing envelope; $t_{\rm C0}$ time corresponding to loss of \(50\%\) of initial cloud mass (transition from Class 0 to Class I); $t_{\rm CI}$ time corresponding to loss of \(90\%\) of initial cloud mass (transition from Class I to Class II); $t_{\rm infall}$ time during which \(99.9\%\) of initial cloud mass $M_{\rm c}$ accretes onto protostar and disk; $t_{\rm d}$ start time of decay of mass infall rate.
    \end{tablenotes}
\end{threeparttable}
\end{table}

\begin{figure}[!ht]
    \centering
    \includegraphics[scale=0.5]{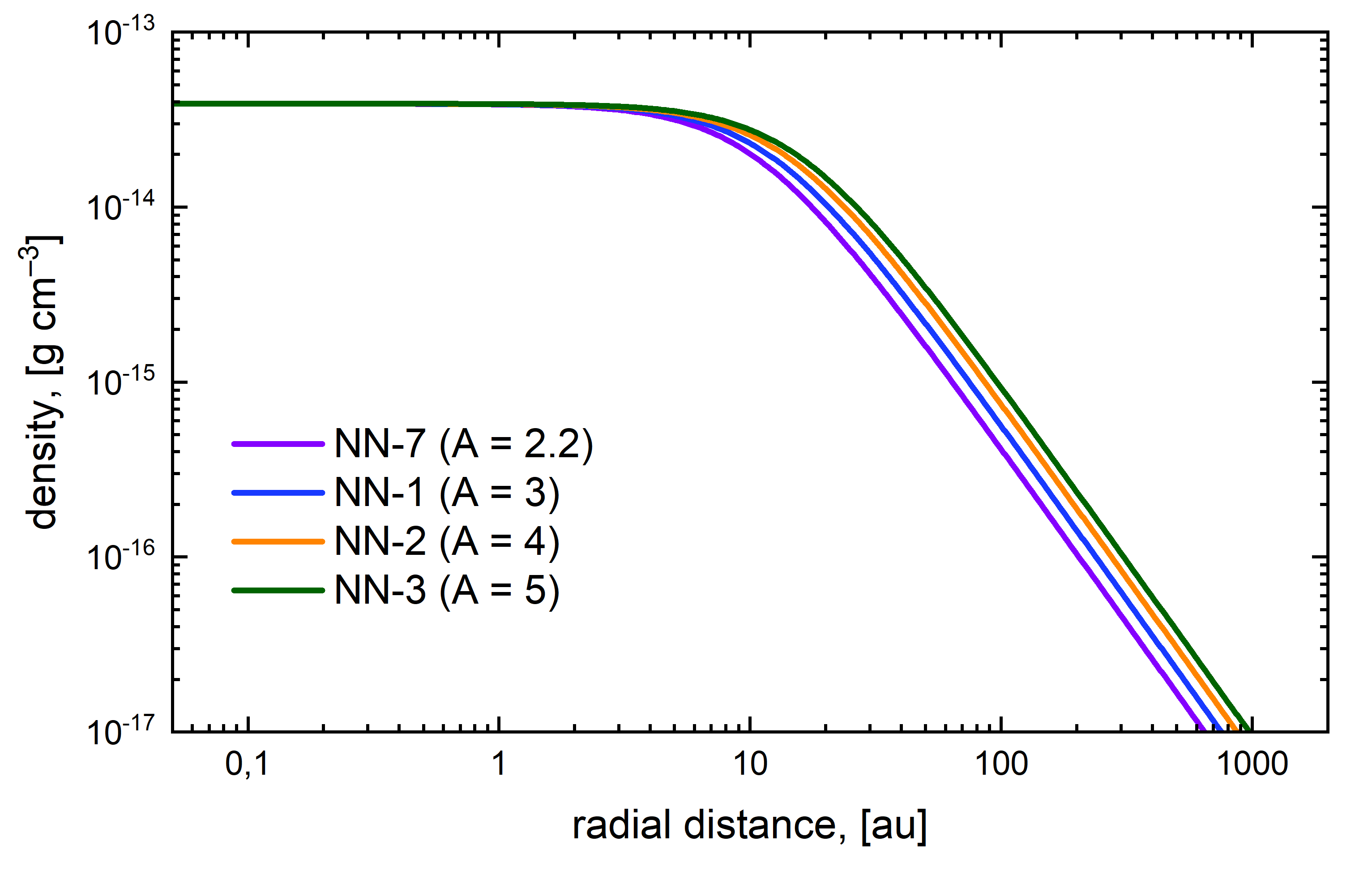}
    \caption{Density distribution $\rho(r)$ in molecular cloud (\ref{rho_c_LP69}) for models NN-1, NN-2, NN-3, and NN-7. For same values of central concentration, mass, and temperature of cloud, it is amplitude of density disturbance $A$ that determines width of central plateau and density at large distances from center of cloud.}
    \label{fig:NNdensity}
\end{figure}

The density distributions in the molecular cloud for models NN-1, NN-2, NN-3 and NN-7 are presented in Fig.~\ref{fig:NNdensity}. All of the listed models use the same values of the central concentration $n_0$, mass $M_{\rm c}$ and temperature $T_{\rm c}$ of the cloud, but the amplitude of the density perturbation $A$ is different. The height of the central plateau depends on $n_0$ (in all models \(10^{10}~\mathrm{cm}^{-3}\)), and the width is determined by the value of $r_0$, which, as can be seen from formulas (\ref{r_0_LP}) and (\ref{factor_k}), increases with the growth of $A$. Therefore, in Fig.~\ref{fig:NNdensity}, for all models, the densities at small distances from the cloud center coincide, and the asymptotic profiles of the models with a large value of $A$ lie higher.

As discussed in Section 2, $A = 2$ corresponds to a density profile that, at large distances from the center of the cloud, is consistent with a marginally stable SIS. Choosing a value of $A > 2$ means that the cloud will be $A/2$ times denser than at equilibrium. Density perturbation and, accordingly, initiation of gravitational collapse can be caused by both internal processes, such as turbulence in a molecular cloud \citep{Ballesteros-Paredes2007, Bate2009}, and external influences, such as interstellar shock waves \citep{Nagakura2009, Boss2012, Cosentino2022} or collisions of molecular clouds \citep{2021PASJ...73S...1F}. The magnitude of the density perturbation may depend on the collapse initiation mechanism, but in this study, $A$ is treated as a free parameter.

\begin{figure}[!ht]
    \centering
    \includegraphics[scale=0.6]{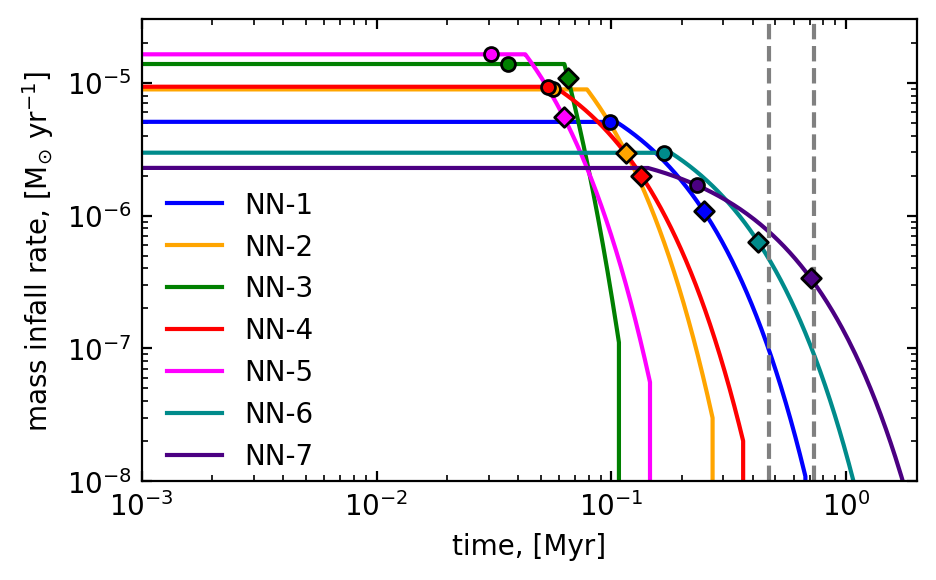}
    \caption{Changes in mass infall rate over time in models NN-1 -- NN-7. Circles show times corresponding to infall of \(50\%\) of initial cloud mass ($t_{\rm C0}$); diamonds show times corresponding to infall of \(90\%\) of mass ($t_{\rm CI}$). Vertical dashed lines indicate boundaries of interval corresponding to most probable total duration of Classes 0 and I according to estimates \citet{Dunham2015}.}
    \label{fig:Mcore-all}
\end{figure}

Fig.~\ref{fig:Mcore-all} shows the mass infall rate onto the disk in the models presented in Table~\ref{tab: table_NNmodels}. In all cases, Eq. (\ref{M_c_dot_M2}) was used for calculations. It is interesting to analyze these rates in the context of the duration of various stages of the evolution of young stellar objects (YSOs). According to \citep{Andre1993} and \citep{Dunham2010}, young Class 0 objects can be defined as systems in which less than \(50\%\) of the initial cloud mass has fallen out. Class I objects correspond to systems in which more than \(50\%\) but less than \(90\%\) of the cloud's mass fell out. The remaining systems can be classified as Class II objects (or T Tauri objects), in which the parent cloud has collapsed almost entirely onto the protostar and disk system.

It can be seen that the loss of \(50\%\) of the mass in most models (except NN-6 and NN-7) occurs in the time interval of up to 100 kyr, while in the NN-3 and NN-5 models, by this time more than \(90\%\) of the mass has already fallen out. However, it is important to note the strong dependence of the results on the cloud temperature $T_{\rm c}$ and amplitude $A$, since it is these values that determine the initial mass infall rate.

At the same time, the observational estimates $t_{\rm C0}$ and $t_{\rm CI}$ differ significantly from the values given in Table~\ref{tab: table_NNmodels}. In \citep{Dunham2015}, intervals corresponding to the most probable duration of 0.15--0.24 Myr for Class 0 and 0.31--0.48 Myr for Class I are presented. As a rule, the time corresponding to the formation of the protostar is chosen as the beginning of the time count. In models NN-1 -- NN-7, time is counted from the onset of gravitational collapse, and the accumulation of mass by the central object sufficient to form a protostar ($M_{\rm star}\approx$ 0.05 $M_{\rm \odot}$) occurs within 10 kyr. From this time, we set aside the previously indicated time intervals, estimated from observations. Therefore, in Fig.~\ref{fig:Mcore-all}, the boundaries of the most probable total duration of Classes 0 and I (or embedded phase) are shown by vertical dashed lines, corresponding to values from 0.47 to 0.73 Myr.

Models NN-6 and NN-7 demonstrate that only the choice of a lower temperature limit ($\approx$ 7~K) or a sufficiently low amplitude $A \approx 2.2$ for the solar mass cloud makes it possible to achieve transition times to Classes I and II comparable with observations. Thus, in the NN-7 model, infall of \(50\%\) of the mass occurs by the time of $t_{\rm C0}$ = 231 kyr, and \(90\%\) by the time of $t_{\rm CI}$ = 706 kyr.

The discrepancy between the modeling results and estimates obtained from observations can be explained by the difficulties associated with distinguishing between YSO Classes. For example, the number of Class 0 objects may be overestimated because edge-on systems appear less evolved due to the high density in the disk plane preventing the passage of the short-wavelength part of the YSO emission spectrum. As a result, the observed Class 0 sources consist of ``true'' Class 0 objects and edge-on Class I objects \citep{Masunaga2000}. The same effect may lead to overestimation of Class II objects compared to Class I objects. Furthermore, accurate observational estimation of the mass of matter in the protostar and disk system, and the envelope is hampered by uncertainties in the opacities, disk temperature, and protostar characteristics \citep{Dunham2014, Vorobyov2017}. The use of alternative methods for estimating the evolutionary phases of YSO, such as bolometric temperature or bolometric luminosity in the submillimeter range, does not always clearly correspond to the definition of evolutionary Classes by mass \citep{Dunham2015}.

Conversely, the discrepancy may also be due to shortcomings of the model used. In particular, the model assumes consideration of an isolated cloud, but recent studies emphasize the importance of taking into account the mass influx from surrounding molecular filaments \citep{Vorobyov2015, Kuffmeier2024}. The introduction of an additional matter into the system may change the rate of deposition of $\dot{M}_{\rm c}$ and, as a result, affect the overall time of $t_{\rm infall}$.

It is also noteworthy that the choice of perturbation amplitude $A = 2$ corresponding to the ``unperturbed'' SIS profile could contribute to a significant increase in the time of matter infall, but in models with the outer boundary calculation and the mass infall rate attenuation, the use of such a value leads to problems associated with unlimited growth of the centrifugal radius $R_{\rm cf}$. As formula (\ref{R_cf_modified}) shows, the centrifugal radius depends on time, and a decrease in the amplitude $A$ to 2 leads to a significant decrease in the initial rate of matter infall and, as a consequence, a significant increase in $t_{\rm infall}$ (thus, with the same $\beta$ values, the infall time in the model with $A = 2$ will be $\approx 1.6$ times greater than in the model with $A = 2.2$). During this time, $R_{\rm cf}$ manages to grow to values exceeding the size of the cloud $r_{\rm out}$, since mathematically the growth of the centrifugal radius is not limited.

\subsection*{3.5. Disk-Mediated Stellar Accretion}

As noted earlier, Eq. (\ref{SigmaInf_NN94}) assumes the matter infall within the entire area bounded by the centrifugal radius. However, one should distinguish between matter infall directly onto the protostar and onto the protoplanetary disk.

It is worth considering separately the case when matter during the collapse process falls only onto the disk, and only then accretes from the disk onto the protostar. This scenario is possible after a protostar forms: due to the initiation of outflows and jets \citep{Arce2007, Lee2020_Jets}, further infall of matter from the cloud directly onto the YSO will be difficult. In this case, almost all of the cloud's matter, with the exception of the mass that formed the protostar at the earliest stages of collapse, will subsequently fall onto the protoplanetary disk.

In this case, it is necessary to make changes to Eq. (\ref{SigmaInf_NN94}). Integral (\ref{CheckNorm_NN94}) should still give the mass $\dot{M}_{\rm c} \Delta t$ that has fallen on the given time layer; however, integration should not occur from zero, but from the inner boundary of the disk $R_{\rm in}$. To take this into account, we introduce into Eq. (\ref{SigmaInf_NN94}) the additional factor $\xi(t)$ which depends only on time and at the selected time layer $\xi = {\rm const}$:
\begin{equation}
\dot{\Sigma}_{\rm infall} (R,t) = \xi \cdot \frac{\dot{M}_{\rm c}}{4\pi R R_{\rm cf}} \left(1 - \frac{R}{R_{\rm cf}} \right)^{-\frac{1}{2}}.
\label{SigmaInf_Inf_Disk_Only}
\end{equation}
Then the integral
\begin{equation}
\int\limits_{R_{\rm in}}^{R_{\rm cf}} \xi \cdot \frac{\dot{M}_{\rm c} \Delta t}{4\pi R R_{\rm cf}} \left(1 - \frac{R}{R_{\rm cf}} \right)^{-\frac{1}{2}} 2\pi R dR = \dot{M}_{\rm c} \Delta t,
\label{Integral_Inf_Disk_Only}
\end{equation}
where
\begin{equation}
\xi = \sqrt{\frac{R_{\rm cf}}{R_{\rm cf} - R_{\rm in}}}.
\label{Xi_Sigma_Dot_Factor}
\end{equation}

Equation (\ref{SigmaInf_Inf_Disk_Only}) with the factor $\xi$ entering into it, which is calculated using formula (\ref{Xi_Sigma_Dot_Factor}), can be considered a more general form of Eq. (\ref{SigmaInf_NN94}), since it allows one to specify the inner boundary from which matter infall will occur. It is easy to see that when $R_{\rm in} = 0$, Eq. (\ref{SigmaInf_Inf_Disk_Only}) transforms into Eq. (\ref{SigmaInf_NN94}).

Note also that Eq. (\ref{SigmaInf_Inf_Disk_Only}) and formula (\ref{Xi_Sigma_Dot_Factor}) come into use only in the case where $R_{\rm cf} > R_{\rm in}$. Until the centrifugal radius reaches the inner boundary of the disk, the mass is transferred directly to the protostar. Table~\ref{tab: table_DiskAccretion} shows the results of calculating the mass that falls out while $R_{\rm cf} \le R_{\rm in}$ = 0.05 au for models NN-1 -- NN-7. Clearly, in each case, a sufficient amount of mass ($\gtrsim$ 0.05 $M_{\rm \odot}$) falls out to form a protostar.

\begin{table}[!ht]
\begin{threeparttable}
    \centering
    \caption{Mass falling directly onto protostar before onset of disk accretion}
    \label{tab: table_DiskAccretion}
    \setlength{\tabcolsep}{14pt}
    \begin{tabularx}{\textwidth}{cccccccc}
    \toprule\noalign{\smallskip}
        \textbf{Model} &  \textbf{NN-1} & \textbf{NN-2} & \textbf{NN-3} & \textbf{NN-4} & \textbf{NN-5} & \textbf{NN-6} &
        \textbf{NN-7} \\ \midrule\noalign{\smallskip}
        \textbf{$M$, $[M_{\rm \odot}]$} & 0.17 & 0.24 & 0.33 & 0.19 & 0.28 & 0.15 & 0.09 \\ \bottomrule
    \end{tabularx}%
\end{threeparttable}
\end{table}

\section*{4. CONCLUSIONS}

The model of matter loss during the gravitational collapse of a molecular cloud, proposed in \citep{Nakamoto&Nakagawa1994}, remains relevant and is used to form a protoplanetary disk in one-dimensional models of viscous evolution. In this article, a number of modifications are proposed aimed at refining this approach while maintaining its computational simplicity. The main modifications are as follows:

\begin{enumerate}
    \item Calculation of the outer boundary of the cloud has been introduced, which makes it possible to determine key parameters of molecular clouds (e.g., the ratio of rotational energy to gravitational energy) that determine the subsequent evolution of protoplanetary disks.
    \item Instead of the idealized distribution of the volume density in a molecular cloud corresponding to a singular isothermal sphere, it is proposed to use a distribution corresponding to a Bonnor-Ebert sphere at the stage of gravitational collapse.
    \item An analytical expression for the rate of mass loss from a collapsing cloud is obtained, which provides for its decay over time, which is justified by the existence of the outer boundary of the cloud.
    \item A case of disk-mediated stellar accretion is described. A more general form of the equation for the rate of mass infall per unit surface area of the protoplanetary disk has been obtained, which provides for specification of the inner boundary of the region within which matter infall will occur after formation of the protostar.
\end{enumerate}

For models including the listed modifications, a study of the duration of the embedded phase was conducted. The simulation results indicate a faster transition of the YSO from Class 0 to Class I and from Class I to Class II compared to observational data, especially in models with a high amplitude of the initial density perturbation. This discrepancy may be due to both the complexity of classifying the observed YSO and the limitations of the original model, which envisages the gravitational collapse of an isolated cloud without an influx of mass from the external medium.

The proposed modifications complement initial model \citep{Nakamoto&Nakagawa1994}, increasing its physical reliability without significantly complicating the approaches used, ensuring the possibility of its use in one-dimensional models and unambiguous interpretation of the results obtained. It is noteworthy that the modifications described in Subsections 3.1 -- 3.3 are also applicable to model \citep{Hueso&Guillot2005} mentioned in the Introduction, since it differs only in its approach to the distribution of matter infall across the disk.

This article is the first in a series of studies on the formation and viscous evolution of the protoplanetary disk. The next paper will present a description of the disk model and the results of numerical simulations of its evolution.

\bibliographystyle{maik.bst}
\bibliography{main.bib}
\clearpage

\section*{Appendix 1}

The equation for the evolution of the surface density in a protoplanetary disk, taking into account the infall of matter from the molecular cloud, can be obtained from the laws of conservation of mass and angular momentum:
\begin{equation}
\frac{\partial \Sigma}{\partial t} + \frac{1}{R} \frac{\partial}{\partial R} \left(R \Sigma v_{\rm R} \right) = \dot{\Sigma}_{\rm infall},
\label{continuity_eq}
\end{equation}
\begin{equation}
\frac{\partial}{\partial t} \left(\Sigma R^3 \Omega_{\rm K} \right) + \frac{\partial}{\partial R} \left(\Sigma \Omega_{\rm K} R^3 v_{\rm R} \right) = \frac{\partial}{\partial R} \left(\Sigma \nu R^3 \frac{\partial \Omega_{\rm K}}{\partial R} \right) + \dot{\Sigma}_{\rm infall} R^3 \Omega,
\label{ang_mom_cons_eq}
\end{equation}
where $v_{\rm R}$ is the radial velocity of gas in the disk.

If we assume $\Omega = \Omega_{\rm K}$ and neglect the self-gravity of the disk, then the expression for the angular velocity of the falling matter will look like
\begin{equation}
\Omega = \sqrt{\frac{GM_{*}}{R^3}},
\label{Omega_Kep}
\end{equation}
where  $M_{*}$ is the mass of the central star. Substituting expression (\ref{Omega_Kep}) and the left-hand side of Eq. (\ref{continuity_eq}) into Eq. (\ref{ang_mom_cons_eq}) instead of $\dot{\Sigma}_{\rm infall}$ yields an expression for the radial velocity:
\begin{equation}
v_{\rm R} = \frac{-3}{\Sigma \sqrt{R}} \frac{\partial}{\partial R} \left(\nu \Sigma \sqrt{R} \right).
\label{v_R_eq}
\end{equation}
Substituting this expression into Eq. (\ref{continuity_eq}), we obtain the classical form of the equation for the evolution of the surface density in the presence of an additional source due to the loss of mass from the envelope:
\begin{equation}
\frac{\partial \Sigma}{\partial t} = \frac{3}{R}\frac{\partial}{\partial R}\left(\sqrt{R} \frac{\partial}{\partial R} \left(\nu \Sigma \sqrt{R} \right)\right)  + \dot{\Sigma}_{\rm infall}.
\label{Pringle_eq_appendix}
\end{equation}

However, according to \citep{Cassen&Moosman1981}, matter infall arrives at the disk at a sub-Keplerian velocity. In this case, equality (\ref{Omega_Kep}) is not satisfied and an additional term will appear in the expression for the radial velocity, due to the decrease in the angular momentum of the disk due to the infall from the envelope:
\begin{equation}
v_{\rm R} = \frac{-3}{\Sigma \sqrt{R}} \frac{\partial}{\partial R} \left(\nu \Sigma \sqrt{R} \right) - \frac{2 R \dot{\Sigma}_{\rm infall}}{\Sigma} \left(1 - \frac{\Omega}{\Omega_{\rm K}} \right).
\label{v_R_eq_new}
\end{equation}
Clearly, for $\Omega < \Omega_{\rm K}$ the difference in angular velocities leads to deceleration of matter in the disk and promotes its movement in the radial direction toward the central star, but in the limiting case, when $\Omega=\Omega_{\rm K}$ expression (\ref{v_R_eq_new}) turns into (\ref{v_R_eq}).

Substituting the new expression for the radial velocity into continuity equation (\ref{continuity_eq}) yields
\begin{equation}
\frac{\partial \Sigma}{\partial t} = \frac{3}{R}\frac{\partial}{\partial R}\left(\sqrt{R} \frac{\partial}{\partial R} \left(\nu \Sigma \sqrt{R} \right)\right)  + \dot{\Sigma}_{\rm infall} + \frac{1}{R} \frac{\partial}{\partial R} \left(2 R^2 \dot{\Sigma}_{\rm infall} \left(1 - \frac{\Omega}{\Omega_{\rm K}} \right) \right).
\label{Pringle_eq_appendix_new}
\end{equation}
Replacing the ratio of angular velocities in accordance to \citep{Cassen&Moosman1981} with
\begin{equation}
\frac{\Omega}{\Omega_{\rm K}} = \frac{v_{\varphi}}{v_{\varphi, \rm K}}=\sqrt{\frac{R}{R_{\rm cf}}},
\label{Omega_to_OmegaKep}
\end{equation}
and also using expression (\ref{SigmaInf_NN94}) for $\dot{\Sigma}_{\rm infall}$ from equality (\ref{Pringle_eq_appendix_new}) the following equation can be obtained for the evolution of the surface density in the protoplanetary disk, taking into account the difference in the angular momentum of the matter infall and matter in the disk:
\begin{equation}
\frac{\partial \Sigma}{\partial t} = \frac{3}{R}\frac{\partial}{\partial R}\left(\sqrt{R} \frac{\partial}{\partial R} \left(\nu \Sigma \sqrt{R} \right)\right)  + \dot{\Sigma}_{\rm infall} + \dot{\Sigma}_{\rm infall} \left(2 - 3 \sqrt{x} + \frac{x}{1 + \sqrt{x}} \right),
\label{Pringle_eq_modified_appendix}
\end{equation}
where $x = R/R_{\rm cf}$ is the ratio of the radial distance $R$ to the centrifugal radius $R_{\rm cf}$.

The use of models of matter infall in which $\Omega \ne \Omega_{\rm K}$, including model \citep{Nakamoto&Nakagawa1994}, assumes the use of Eq. (\ref{Pringle_eq_modified_appendix}), not (\ref{Pringle_eq_appendix}).

\section*{Appendix 2}

According to \citep{Kratter2010}, the coefficient $m_0$ included in Eq. (\ref{Mdotcore_Shu77}) is related to the amplitude of the density disturbance $A$ by
\begin{equation}
    m_0 = \left\{
    \begin{alignedat}{2} 
        0.975, & \quad A = 2, \\
        \frac{(2A)^{\frac{3}{2}}}{\pi}, & \quad A \gg 2.
    \end{alignedat}
    \right.
\label{m0_Kratter}
\end{equation}
Table~\ref{tab: table_mA_relation} shows the values of $m_0^{\rm Sh}$ depending on the amplitude $A$, calculated using polynomial (\ref{polynom_ShuTable}), the values of $m_0^{\rm Kr}$ for the same $A$, calculated using formula (\ref{m0_Kratter}) from work \citep{Kratter2010}, as well as their ratio $m_0^{\rm Kr}/{m_0^{\rm Sh}}$. As the table shows, the values of $m_0$ calculated using formula (\ref{m0_Kratter}) are overestimated for $2 < A \lesssim 5$ compared to those given in \citep{Shu1977} or calculated using polynomial (\ref{polynom_ShuTable}). Since the rate of matter infall from the cloud is proportional to $m_0$, a difference of even 1.5--1.2 times can significantly affect the time of matter infall, especially in cases where decay in $\dot{M}_{\rm c}$ due to the presence of the outer boundary of the cloud is taken into account. Therefore, for relatively small $A$, one should use polynomial (\ref{polynom_ShuTable}) rather than formula (\ref{m0_Kratter}).

\begin{table}[!ht]
\begin{threeparttable}
    \centering
    \caption{Comparison of values $m_0$ calculated using formulas (\ref{polynom_ShuTable}) and (\ref{m0_Kratter}).}
    \label{tab: table_mA_relation}
    \footnotesize
    \setlength{\tabcolsep}{2.2pt}
    \scriptsize
    \setlength{\tabcolsep}{2.2pt}
    \renewcommand{\arraystretch}{1.1}
    \begin{tabular}{|c|*{21}{c|}}
    \hline
        \textbf{$A$} &
         2.2 & 2.4 & 2.6 & 2.8 & 3.0 & 3.2 & 3.4 & 3.6 & 3.8 & 4.0 & 4.2 & 4.4 & 4.6 & 4.8 & 5.0 & 5.2 & 5.4 & 5.6 & 5.8 & 6.0 \\
    \hline
        \textbf{$m_0^{\rm Sh}$} & 1.43 & 1.87 & 2.31 & 2.75 & 3.19 & 3.64 & 4.10 & 4.57 & 5.07 & 5.59 & 6.14 & 6.72 & 7.33 & 7.98 & 8.68 & 9.42 & 10.22 & 11.07 & 11.97 & 12.94 \\
    \hline
        \textbf{$m_0^{\rm Kr}$} & 2.94 & 3.35 & 3.77 & 4.22 & 4.68 & 5.15 & 5.64 & 6.15 & 6.67 & 7.20 & 7.75 & 8.31 & 8.88 & 9.47 & 10.07 & 10.68 & 11.30 & 11.93 & 12.58 & 13.23 \\
    \hline
        \textbf{$m_0^{\rm Kr}/m_0^{\rm Sh}$} & 2.05 & 1.79 & 1.63 & 1.54 & 1.47 & 1.42 & 1.38 & 1.34 & 1.32 & 1.29 & 1.26 & 1.24 & 1.21 & 1.19 & 1.16 & 1.13 & 1.11 & 1.08 & 1.05 & 1.02 \\
    \hline
    \end{tabular}%
\end{threeparttable}
\end{table}

\section*{ACKNOWLEDGMENTS}

The authors of the article thank the anonymous reviewer for useful advice and recommendations, which contributed to significant improvement of the article.

\section*{FUNDING}

The study was supported by the Ministry of Science and Higher Education of the Russian Federation, state assignment no. GZ0110/23-10-IF.

\end{document}